\documentclass[runningheads]{llncs}
\usepackage{graphicx}
\usepackage{xcolor}
\usepackage{amsmath}
\usepackage{amssymb}
\usepackage{multirow}
\usepackage{booktabs}
\usepackage{hyperref}
\begin{document}
\title{
Estimating Neural Orientation Distribution Fields on High Resolution Diffusion MRI Scans
}
\titlerunning{Neural Orientation Distribution Fields on High-Res Diffusion MRI Scans}
\author{
Mohammed Munzer Dwedari*\inst{1,2} \and
William Consagra*\inst{1} \and
Philip Müller\inst{2} \and
Özgün Turgut\inst{2} \and
Daniel Rueckert\inst{2} \and
Yogesh Rathi\inst{1}
}%index{Dwedari, Mohammed Munzer}%index{Consagra, William}%index{Müller, Philip}%index{Turgut, Özgün}%index{Rueckert, Daniel}%index{Rathi, Yogesh}

\def\thefootnote{*}\footnotetext{Authors contributed equally to this work}\def\thefootnote{\arabic{footnote}}
\authorrunning{Dwedari et al.}
\institute{
Psychiatry Neuroimaging Laboratory, Brigham and Women’s Hospital, Harvard Medical School, Boston, USA
\email{\{mdwedari,wconsagra,yogesh\}@bwh.harvard.edu}
\and
Technical University of Munich, Munich, Germany
\email{\{munzer.dwedari,philip.j.mueller,oezguen.turgut,daniel.rueckert\}@tum.de}
}

\maketitle              % typeset the header of the contribution
%

% ABSTRACT
\begin{abstract}
The Orientation Distribution Function (ODF) characterizes key brain microstructural properties and plays an important role in understanding brain structural connectivity.
Recent works introduced Implicit Neural Representation (INR) based approaches to form a spatially aware continuous estimate of the ODF field and demonstrated promising results in key tasks of interest when compared to conventional discrete approaches.
However, traditional INR methods face difficulties when scaling to large-scale images, such as modern ultra-high-resolution MRI scans, posing challenges in learning fine structures as well as inefficiencies in training and inference speed. In this work, we propose \emph{HashEnc}, a grid-hash-encoding-based estimation of the ODF field and demonstrate its effectiveness in retaining structural and textural features. We show that HashEnc achieves a 10$\,$\% enhancement in image quality while requiring 3x less computational resources than current methods. Our code can be found at \href{https://github.com/MunzerDw/NODF-HashEnc}{https://github.com/MunzerDw/NODF-HashEnc}.

\end{abstract}

% INTRODUCTION
\section{Introduction}

The Orientation Distribution Function (ODF) plays an important role for understanding brain structural connectivity and brain-based disorders \cite{tuch2004}. It describes the angular probability distribution of water molecule diffusion in brain tissue \cite{estodf_2021}, where water diffusion is strengthened along the direction of white matter fiber tracts. Thus, the ODF serves as an indirect characterization of the white matter fiber structure at a given voxel in the brain, which provides critical information for tractography and microstructure estimation \cite{tractogrophy}.

Consagra \textit{et al.} \cite{nodf} introduced an Implicit Neural Representation (INR) framework (called NODF), using Sinusoidal Representation Networks (SIREN) \cite{siren}, to model a spatially aware continuous ODF field, estimated from diffusion signal. Their method enables resolution-agnostic estimation and uncertainty quantification of the ODFs. However, while large SIRENs can estimate the ODF field on individual slices and regions of interest, their ability to learn a continuous field for images on the scale of modern high-resolution whole brain scans \cite{highresdata} suffers from days-long training times \cite{kilonerf}, rendering them impractical for these applications. This issue arises because all the network weights need to be evaluated and updated during each pass. Moreover, the fine-tuning of important hyperparameters, including those for regularization and the sine frequency, becomes difficult or even computationally infeasible due to repeated network training.

To mitigate these issues, we investigate a solution, referred to as \emph{HashEnc}, based on the grid-like local embeddings as proposed by Müller \textit{et al.} \cite{instantngb} to replace SIREN in the NODF framework. The use of grid-like embeddings allows HashEnc to store local information of the training subject. As every region has designated embeddings, only the local neighborhood embeddings and the small MLP weights need to be updated during training. Therefore, the required MLP size to predict the final output signal for a coordinate becomes much smaller and thus efficient to train. 
We train HashEnc on a submillimeter resolution and low signal-to-noise-ratio diffusion MRI (dMRI) scan \cite{highresdata} and evaluate the results on highly detailed areas such as the cerebellum. While SIREN tends to over-smooth the estimated ODF, we demonstrate the capability of HashEnc to learn fine structural and textural details in significantly less training time. In summary, our contributions are:
\begin{itemize}
\item We propose \emph{HashEnc}, a grid-hash-encoding-based INR that represents a ``field" of ODFs in a spatially continuous manner across any ultra-high-resolution dMRI scan.
\item We quantitatively and qualitatively compare HashEnc with SIREN, where HashEnc achieves a 10\% enhancement in image quality while being up to 3x faster to train.
\item We study the key characteristics of HashEnc through ablation studies.
\end{itemize}

% RELATED WORK
\section{Related Work}
\subsubsection{Orientation Distribution Function.} The estimation of ODFs from diffusion signals poses a challenging inverse problem, that has mostly been tackled voxel-wise \cite{descoteaux2007,sphridgelets} or by incorporating neighborhood information \cite{adaptivesmoothing,mspoassmoothing}. Furthermore, other lines of work introduce machine learning techniques to directly estimate ODFs from diffusion signal through supervised or unsupervised training \cite{aeodfs,resdnn,spears2023learning}. Recently, \cite{nodf} utilized INRs to continuously parameterize the ODF field and derive a conditional posterior distribution for uncertainty quantification.

\subsubsection{Implicit Neural Representations in Medical Imaging.} INRs are increasingly utilized in computer vision and medical imaging, enabling continuous modeling of discrete data with minimal memory usage \cite{nerf_2021,inrocc_2019,meshfreeflownet,pifu,pixelnerf,deepsdf,inrmedsurvey,hendriks2023neural}. Their flexibility and differentiability facilitate various tasks, such as image reconstruction \cite{ArSSR,nesvor,spears2023learning,ewert2024geometric}, segmentation \cite{nerd,retinalseg,binaryseg}, and registration \cite{IDRI,mirnf,inrbrainregist}, effectively addressing issues like scarce data and lengthy acquisition times. INRs are also used for inverse imaging tasks \cite{dynamicctinr,nodf,coil} or 3D volume reconstruction from sparse 2D images \cite{Mednerf,nerf3dxray}.

% METHOD
\section{Method}

\subsection{Background and Notation}
\subsubsection{Orientation Distribution Function.} The Orientation Distribution Function (ODF) at a voxel \( \boldsymbol{v} \), \( g(\boldsymbol{v},\cdot) \), describes the angular distribution of water molecule diffusion and is connected to the diffusion signal, denoted as $f(\boldsymbol{v},\cdot)\mapsto\mathbb{R}^{+}$, by the Funk-Radon transform (FRT). To compute the ODF, we truncate the spherical harmonic basis \cite{tuch2004} at a finite rank \( K \), modeling it as:
\begin{equation}
\label{eqn:odf}
    g(\boldsymbol{v},\boldsymbol{p}) = \sum_{k=1}^K c_{k}(\boldsymbol{v})\phi_{k}(\boldsymbol{p}),
\end{equation}
where \(\phi_{k}(\boldsymbol{p})\) are the harmonic basis functions and \( c_{k}(\boldsymbol{v}) \) are the harmonic coefficients. Measurements on \( M \) spherical locations $\boldsymbol{p_1},....,\boldsymbol{p_M}$, called gradient directions, at each of a regular set of voxel locations $\boldsymbol{v}_1, ..., \boldsymbol{v}_{N}$, translate to a noisy Gaussian model linking observed signals to the coefficients \( c_{k}(\boldsymbol{v}) \):
\begin{equation}\label{eqn:measurement_model}
    \boldsymbol{y}_{i} := (y_{i,1},...,y_{i,M}) \sim \mathcal{N}(\boldsymbol{\Phi}\boldsymbol{G}\boldsymbol{c}(\boldsymbol{v}_i), \sigma_e^2\boldsymbol{I}_{M}),
\end{equation}
where $\boldsymbol{c}(\boldsymbol{v})=(c_{1}(\boldsymbol{v}), ..., c_{K}(\boldsymbol{v}))^{\intercal}$, $\boldsymbol{G}\in\mathbb{R}^{K\times K}$ is the diagonal inverse matrix of the FRT, $\boldsymbol{\Phi} \in \mathbb{R}^{M \times K}$ is the evaluation of the $K$ real-symmetric spherical harmonic basis functions along all $M$ gradient directions, $\boldsymbol{I}_{M}$ is the $M$-dimensional identity matrix, and $\sigma_e^2$ is the measurement error variance. 

\subsubsection{Neural Orientation Distribution Fields.} The NODF framework introduces an implicit model to capture the spatial correlation in the ODF field through a rank \( r \) spatial basis learned via an implicit neural representation \( \boldsymbol{\xi}_{\boldsymbol{\theta}}:\mathbb{R}^{3}\mapsto\mathbb{R}^r \). This basis is used to construct the harmonic coefficient fields via a multivariate linear basis expansion \( \boldsymbol{c}(\boldsymbol{v})=\boldsymbol{W}\boldsymbol{\xi}_{\boldsymbol{\theta}}(\boldsymbol{v}) \), with \( \boldsymbol{W} \) being a matrix in \( \mathbb{R}^{K\times r} \).

Following the framework, with a matrix normal prior on \( \boldsymbol{W} \) inducing a Gaussian process prior on \( g(\boldsymbol{v},\cdot) \), and given the normal likelihood \eqref{eqn:measurement_model}, the posterior distribution can be derived as given in \cite{nodf}:
\begin{equation}
\label{eqn:weight_posterior}
\text{vec}(\boldsymbol{W}) | \boldsymbol{V}, \boldsymbol{Y}, \boldsymbol{\theta}, \gamma, \sigma_w^2,  \sigma_e^2 \sim \mathcal{N}_{Kr}\left(\frac{1}{\sigma_{e}^2}\boldsymbol{\Lambda}_{\boldsymbol{\theta}}^{-1}[\boldsymbol{\Xi}_{\boldsymbol{\theta}}^{\intercal}\otimes\boldsymbol{\Phi}\boldsymbol{G}]^{\intercal}\text{vec}(\boldsymbol{Y}), \boldsymbol{\Lambda}_{\boldsymbol{\theta}}^{-1}\right),
\end{equation}
\begin{equation}
    \boldsymbol{\Lambda}_{\boldsymbol{\theta}} = \frac{1}{\sigma^2_{e}}\left(\frac{\sigma^2_{e}}{\sigma^2_{w}}\boldsymbol{I}_{r}\otimes\boldsymbol{R}_{\gamma} +  \boldsymbol{\Xi}_{\boldsymbol{\theta}}\boldsymbol{\Xi}_{\boldsymbol{\theta}}^{\intercal}\otimes[\boldsymbol{\Phi}\boldsymbol{G}]^{\intercal}\boldsymbol{\Phi}\boldsymbol{G}\right)
\end{equation}
where \(\text{vec}\) is the vectorization operator, \(\otimes\) denotes the Kronecker product, and \( \boldsymbol{R}_{\gamma} \) is the covariance matrix from a spherical Matern Gaussian process with parameters \( \gamma \), $\boldsymbol{Y} = [\boldsymbol{y}_1^{\intercal}, ..., \boldsymbol{y}_{N}^{\intercal}] \in \mathbb{R}^{M\times N}$, $\boldsymbol{\Xi}_{\boldsymbol{\theta}} = [\boldsymbol{\xi}_{\boldsymbol{\theta}}^{\intercal}(\boldsymbol{v}_1),...,\boldsymbol{\xi}_{\boldsymbol{\theta}}^{\intercal}(\boldsymbol{v}_N)]^{\intercal}\in\mathbb{R}^{r\times N}$, and $\boldsymbol{V}=[\boldsymbol{v}_{1},...,\boldsymbol{v}_N] \in \mathbb{R}^{N\times 3}$. The unknown conditioning parameters of \eqref{eqn:weight_posterior} are then estimated and plugged in for inference, i.e. point estimation and uncertainty quantification. Specifically, the network parameters $\widehat{\boldsymbol{\theta}}$ are estimated using stochastic gradient descent on a regularized variant of the negative log likelihood, where the regularization strength $\lambda_c$ is selected using a Bayesian optimization scheme. We follow the same procedure as in \cite{nodf} to estimate the remaining variance parameters $\sigma_{e}^2,\sigma_{w}^2$. When estimating a quantity of interest (QOI) from the ODFs for downstream tasks, e.g. fractional anisotropy or principal diffusion directions, we quantify the uncertainty in the QOI by sampling the ODF field (through the posterior \eqref{eqn:weight_posterior}) and determining a confidence interval.

\subsection{Grid-Hash-Encoding of Harmonic Coefficient Fields}\label{ssec:HGE}

\begin{figure}
    \centering
    \includegraphics[scale=0.36]{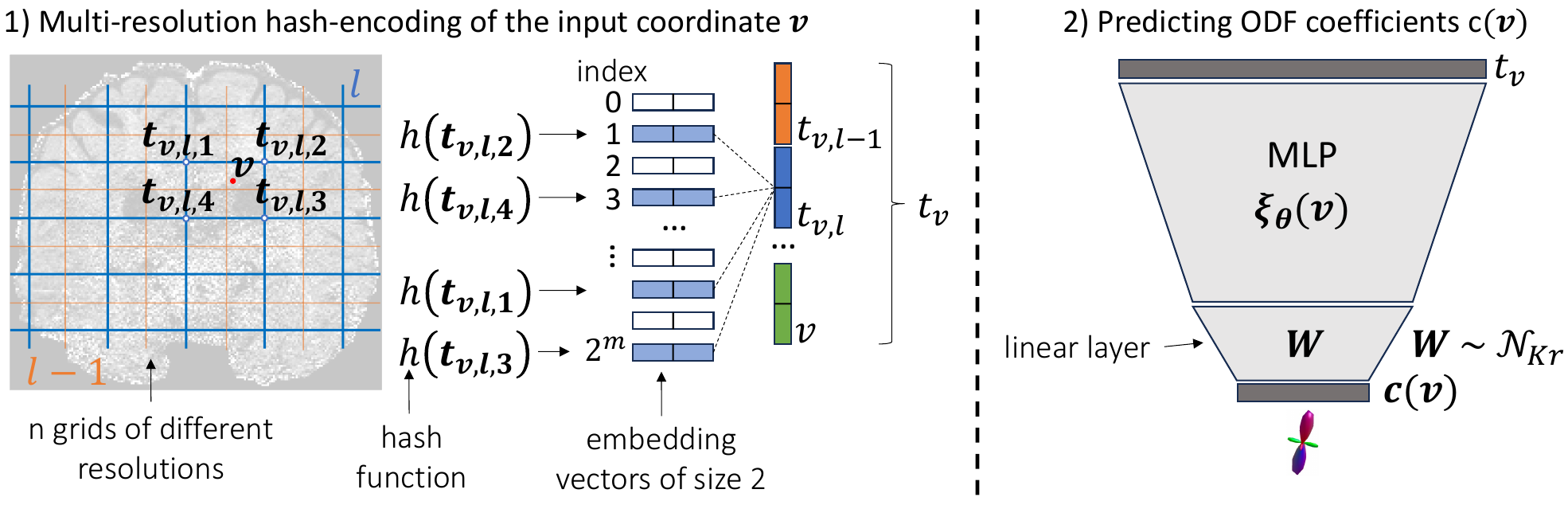}
    \caption{Overview of \emph{HashEnc}. 1) Given point $\boldsymbol{v}$, for each resolution grid $l$, embedding vectors of the surrounding corner points are retrieved from the lookup table by hashing their grid coordinates $t_{v,l,i}$. Then, the corner embeddings are combined into one vector via linear interpolation. The final embedding vector $t_v$ is obtained by concatenating the input coordinates $\boldsymbol{v}$ and other grid vectors $t_{v,l}$. The grid is shown in 2D instead of 3D for the sake of clarity. 2) $t_v$ is fed into a SIREN and processed by a linear layer $W$ to output the spherical harmonic coefficients $\boldsymbol{c}(\boldsymbol{v})$.}
    \label{fig:method}
\end{figure}

Using a single SIREN MLP for spatial basis \( \boldsymbol{\xi}_{\boldsymbol{\theta}} \) in large images leads to computational challenges. Due to the need for a large network rank \( r \), the inversion of \( \boldsymbol{\Lambda}_{\boldsymbol{\theta}} \) can become complex and unstable. In addition, gradient computation for \( \boldsymbol{\theta} \) during backpropagation is slow, as it requires evaluating all parameters for every voxel. To address this, we propose adopting a grid-hash-encoding method (HashEnc) \cite{instantngb} with a much smaller MLP that trains more quickly, leveraging local embedding vectors for storing regional information.

The input of the network is a 3D coordinate $\boldsymbol{v} \in \mathbb{R}^3$ and the output is the real-symmetric spherical harmonic expansion coefficients of the ODF $\boldsymbol{c}(\boldsymbol{v}) \in \mathbb{R}^K$, where $K=45$ (Figure \ref{fig:method}). Each resolution grid $l$ takes the 3D coordinates as input and retrieves the grid coordinates of the 8 surrounding corners. These surrounding coordinates are hashed into index values and the corresponding embedding vectors are retrieved from a dictionary of size $2^m$ belonging to that grid. Via linear interpolation, at each resolution grid $l$ the embedding vectors are interpolated into one vector $t_{v,l}$. The vectors of all $n$ resolution grids are then concatenated together into one vector along with the input coordinates, forming $t_v = (t_{v,1},...,t_{v,n},\boldsymbol{v})$. $t_{\boldsymbol{v}}$ is then passed into an MLP head (SIREN of 2x64 hidden layers) to predict the spatial basis $\xi_{\boldsymbol{\theta}}(\boldsymbol{v})$, which is subsequently multiplied with the linear layer $\boldsymbol{W}$ to obtain the $K$ ODF coefficients $\boldsymbol{c}(\boldsymbol{v})$. Using the inverse of the Funk-Radon transform $\boldsymbol{G}$, we obtain the coefficients of the signal $f$ expansion over the real-symmetric spherical harmonic basis. From the $\boldsymbol{p}$ points on the sphere, indicating the $M$ gradient directions, we can obtain the diffusion signals $f(\boldsymbol{v},\boldsymbol{p}_1), ...,f(\boldsymbol{v},\boldsymbol{p}_M)$.

% To illustrate the computational advantage: for inference at a single voxel, HashEnc uses ~13,000 parameters (31 input size, 2x 64 hidden, 45 output size), outpacing SIREN's millions of parameters. Doubling HashEnc's capacity from 2 to 4 embedding vector size adds $<$2,000 parameters (59 input size), while doubling SIREN's requires millions more per voxel, significantly increasing computation time.

To illustrate the computational advantage: for inference at a single voxel, HashEnc uses ~13,000 parameters (31 input, 2x 64 hidden, 45 output size), $<$0.1\% of the total number of parameters.
In contrast, SIREN requires all its parameters for each voxel. Doubling HashEnc's capacity from 2 to 4 embedding vector size adds $<$2,000 parameters (59 input size), while doubling SIREN's doubles the parameters required per voxel, leading to significant computational challenges.

% To illustrate the speed advantage: for inference at a single voxel, HashEnc requires around 13,000 parameters (31 input vector size, 2x 64-sized hidden layers, 45 output vector size), speeding up both forward and backward passes compared to a single large SIREN MLP with millions of parameters (see section \ref{sec:experimental_setup}). Increasing HashEnc’s embedding vectors from 2 to 4 doubles its capacity but adds less than 2 thousand parameters to the forward/backward pass (input vector size becomes 59). In contrast, doubling SIREN’s parameters means that millions additional parameters are required for each voxel inference, greatly increasing computation time.

% EXPERIMENTAL SETUP
\section{Experimental Setup}
\label{sec:experimental_setup}

\label{sec:data}
\subsubsection{Data.} We train on the publicly available high-resolution data (760 $\mu m^3$) from \cite{highresdata}\footnote{License: \url{http://creativecommons.org/licenses/by/4.0/}}. The data consists of multiple scan sessions with 420 gradient directions at $b=1,000s/mm^2$. We train on one of the scan session data (with 70 gradient directions) which has very low SNR. As there is no ground truth data, we consider a 6 session average (across 420 directions) as a reasonable ground truth image and apply penalized Spherical Harmonics Least Square (SHLS) from \cite{descoteaux2007} to derive ground truth ODFs. The dimension of the resulting image is $190\times224\times178\times M$.

\subsubsection{Training and Evaluation.} We compare HashEnc and SIREN, the latter being an MLP with 10 layers of 1024 units each and sine activations. SIREN is trained with a learning rate of 1e-6 for 10,000 epochs. We use Bayesian optimization on a single slice to select $\lambda_c$. While potentially sub-optimal, this approach is computationally necessary, especially for SIREN, as full volume training takes days. HashEnc employs 14 resolution grid levels, starting from resolution size 6, with a $2^{20}$-sized lookup table per level. Both methods are trained on an RTX 4090 GPU with $M=70$, $M=40$, and $M=20$ gradient directions using PyTorch.

We evaluate both methods on the Feature Similarity Index (FSIM)\footnote{Calculated via: \url{https://pypi.org/project/image-similarity-measures/}} \cite{fsim}, which mimics the human perception to images and focuses on features such as edges, corners, and textures. It consists of two components: Phase Congruency PC and Gradient Magnitude (GM). PC compares feature points regardless of brightness or contrast. GM captures the image edge information by measuring the gradient magnitude of the image. This constellation is important for distinguishing the tendency of SIREN to over-smooth and HashEnc to overfit noise. FSIM scores for gray scale General Fractional Anisotropy (GFA) and RGB Diffusion Tensor (DTI) images across all sagittal, axial and coronal slices are calculated against the 6-session average, with median values reported in Table \ref{tab:quant_results} and sample images provided in the supplementary material. GFA calculates the degree of anisotropy of water diffusion at each voxel, where a higher value indicates stronger anisotropy. The Diffusion Tensor Image (DTI) indicates via RGB coloring the dominant fiber orientation direction at each voxel.

Furthermore, GFA images are visualized in Figure \ref{fig:time_comparison}. DTI images and deconvolved ODFs (using Constrained Spherical Deconvolution \cite{csd}) on a small sagittal section in the Cerebellum are shown in Figure \ref{fig:qualitative_results}. To quantify the uncertainty of each method, the posterior is sampled 250 times. Then, the voxel-wise GFA is determined for each sampled ODF field. The uncertainty on the GFA is analysed via the standard deviation to mean ratio.

% RESULTS
\section{Results}
\label{sec:results}

\subsection{Comparison with Current Methods}

\begin{table}
\caption{Median value of Feature Similarity Index (FSIM) to the 6 session average of all sagittal, axial and coronal slices. FSIM-GFA is calculated on gray scale GFA images, and FSIM-DTI is calculated on RGB DTI images. 1 means perfect similarity. Visuals are provided in the supplementary materials. HashEnc performs better in all settings except for $M=20$ on FSIM-DTI.}
\centering
% \resizebox{\textwidth}
% {!}{%
\setlength{\tabcolsep}{3pt}
\begin{tabular}{ccccc}
\toprule
Gradient Directions ($M$) & Model & FSIM-GFA & FSIM-DTI \\ \midrule
\multirow{2}{*}{70} & SIREN    & 0.55 & 0.61 \\ 
 & HashEnc  & \textbf{0.66} & \textbf{0.68} \\  \midrule
\multirow{2}{*}{40} & SIREN    & 0.62 & 0.63 \\ 
  & HashEnc  & \textbf{0.66} & \textbf{0.66} \\  \midrule
\multirow{2}{*}{20} & SIREN    & 0.62 & \textbf{0.64} \\ 
 & HashEnc  & \textbf{0.64} & 0.62 \\ \bottomrule
\end{tabular}
% %
% }
\label{tab:quant_results}
\end{table}

\begin{figure}
    \centering
    \includegraphics[scale=0.29]{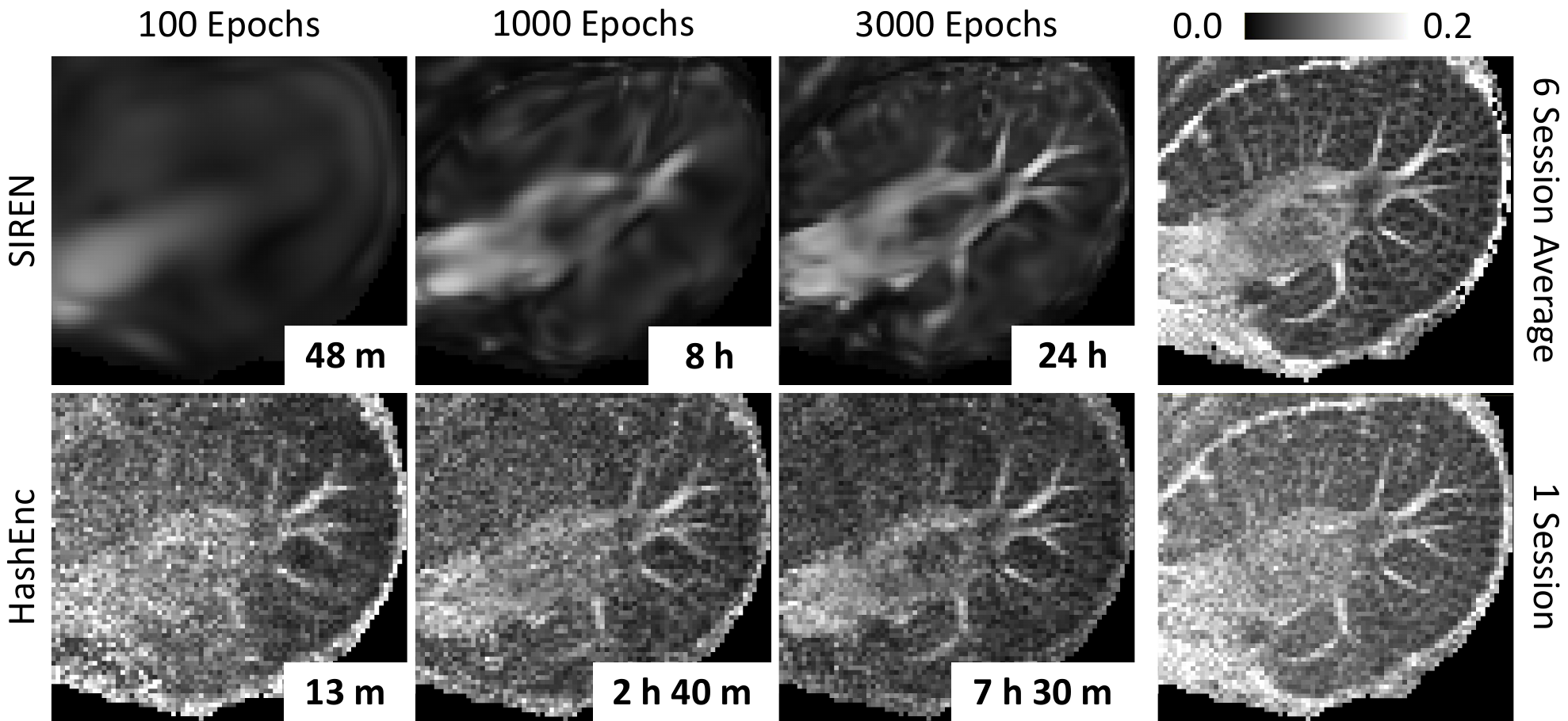}
    \caption{GFA reconstruction images from a sagittal cerebellum slice at different training times with $M=70$ gradient directions. The scale on the top right indicates the degree of anisotropy of water diffusion. Included are GFA of the training image (1 session) and the 6 session average for reference. HashEnc fits a much more detailed ODF field after significantly less training time compared to SIREN.}
    \label{fig:time_comparison}
\end{figure}

\begin{figure}[ht]
    \centering
    \includegraphics[scale=0.35]{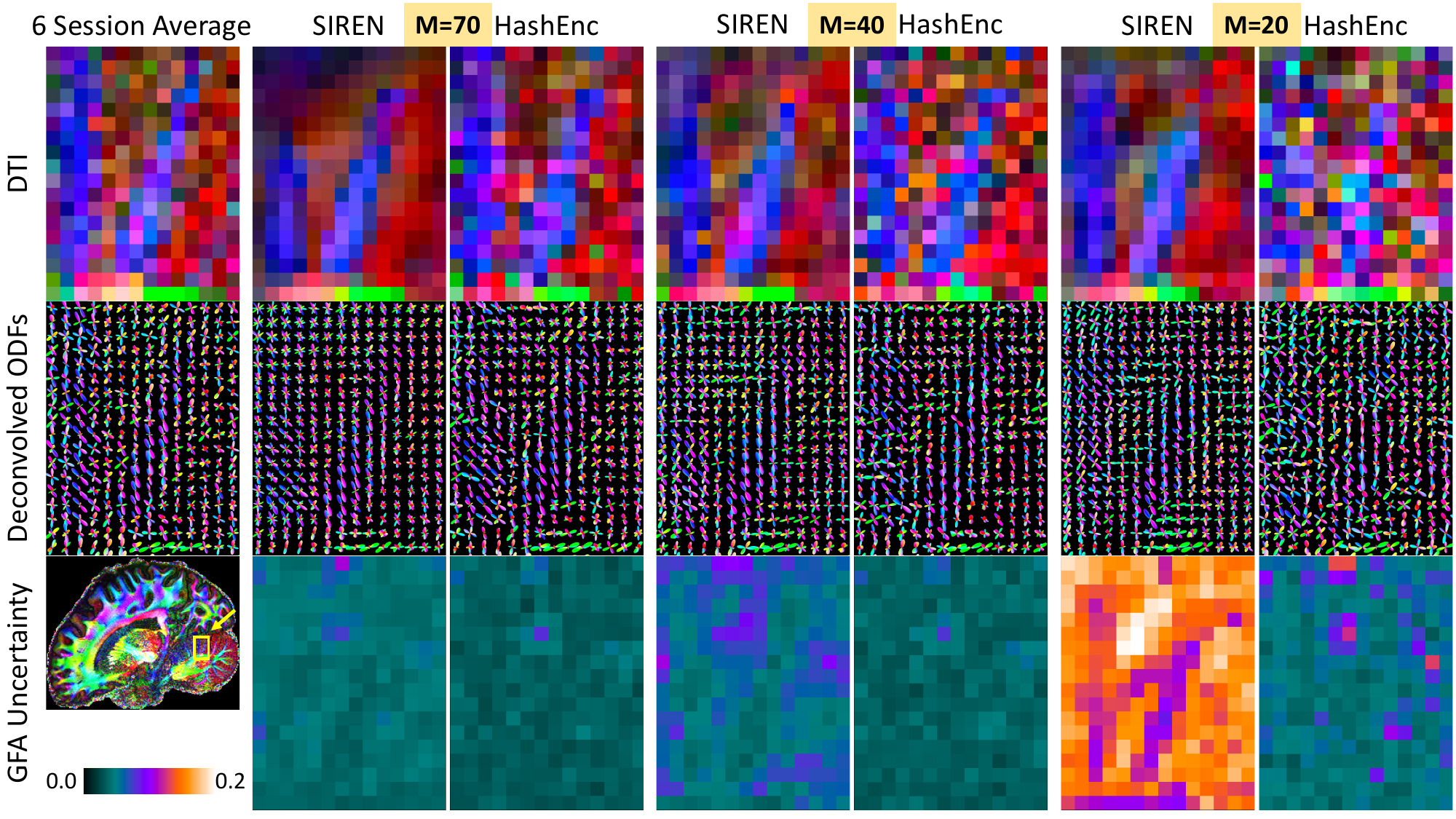}
    \caption{Qualitative reconstruction examples on a small sagittal Cerebellum section, showcasing DTI images, deconvolved ODFs, and GFA uncertainty. The scale on the bottom left indicates variability in the GFA of the ODF samples. SIREN and HashEnc are trained for 10,000 and 3,000 epochs, respectively, and compared for $M=[70, 40, 20]$ gradient directions. SIREN tends to over-smooth the ODF field, particularly at $M=70$, but is more robust with fewer gradient directions. HashEnc matches the structural and textural details of the 6-session average better and exhibits less uncertainty.}
    \label{fig:qualitative_results}
\end{figure}

Compared to SIREN, HashEnc shows a higher structural similarity to the 6 session average volume in terms of GFA gray scale and DTI RGB images. This is especially apparent for $M=70$, where SIREN shows worse performance to $M=40$ and $M=20$ (Table \ref{tab:quant_results}). One reason for the lower similarity scores of SIREN is the over-smoothing effect that SIREN shows, which creates blurry spots in the image (see Figures \ref{fig:time_comparison} and \ref{fig:qualitative_results}). We note that this over-smoothing effect in SIREN might be mitigated by globally tuning its hyperparameters, i.e., selecting $\lambda_c$ by training on the whole image rather than just a slice. However, automatically doing this through Bayesian Optimization is highly computationally problematic due to the excessively long training times. This further underscores the advantage of HashEnc's faster training times. On the other hand, SIREN shows a robust performance as $M$ gets lower and outperforms HashEnc on FSIM-DTI for $M=20$. Further visuals are provided in the supplementary material.

The dominant advantage of HashEnc are the training and inference times, allowing for much faster estimation of the ODF field and evaluation for downstream tasks. HashEnc can already produce fine-grained estimations after 100 epochs of training, which is not the case for SIREN (Figure \ref{fig:time_comparison}). After 1000 training epochs, HashEnc shows similar but more detailed results whereas SIREN is yet to fit fine-grained regions. The time efficiency comes from using multi-resolution grid embeddings that store local volume information at various resolution levels, allowing for a smaller MLP head and the use of a larger learning rate. As for inference speed, HashEnc requires about 4 seconds to infer the ODF coefficients on the whole brain while SIREN requires 173 seconds (tested on CPU).
When SIREN is trained for 10,000 epochs, visually, both methods produce results with comparable performance but different characteristics (Figure \ref{fig:qualitative_results}). SIREN produces overly-smooth estimates of the ODF field, resulting in slightly blurry but less noisy look, as demonstrated in Figure \ref{fig:qualitative_results}. This can be observed specifically in areas with high contrast, such as between white and gray matter. The gradual transition of SIREN, which is visible on the borders of the blue fiber tracts in the ODF images for $M=70$ and $M=40$, is also reflected in the low FSIM scores in Table \ref{tab:quant_results}. HashEnc on the other hand learns individual details better in fine-grained regions as can be seen in the width of the blue fiber tracts in the DTI images of $M=70$ and $M=40$. However, it tends to overfit to noise easier, such as in the cases of $M=40$ and $M=20$. For these (lower) gradient directions, SIREN is more robust to noise. As for uncertainty quantification, HashEnc shows a consistent and lower uncertainty, whereas SIREN exhibits larger uncertainty especially in the border regions between white and gray matter at $M=40$ and $M=20$ gradients. Additionally, HashEnc computes posterior \eqref{eqn:weight_posterior} means and variances faster due to its smaller $\boldsymbol{W}$ matrix size.

\subsection{Ablation Studies}
\subsubsection{How do grid resolution levels and lookup table size affect the characteristics of the ODF field?} Different resolution levels $n$ (12$-$14) and lookup table sizes $2^m$ ($m=19$ and $m=20$) are analysed. A longer and thinner estimation of the fiber tracts can be observed for $m=20$ (see Figure 2 in supplementary material). On the other hand, having a higher number of resolution levels shows finer but noisier details, whereas for $n=12$ resolution levels the image looks smoother with some information lost (e.g. the tip of the thin blue fiber tracts). Quantitatively, $n=14$ resolution levels and $2^{20}$ lookup table size shows the highest feature similarity score to the 6 session average.

\subsubsection{How does the MLP head affect HashEnc?} In this experiment, we try three types of MLP heads, including SIREN \cite{siren}, WIRE \cite{wire}, and ReLU \cite{instantngb}. Our experiments show that there is no significant difference both visually (DTI images) and quantitatively (FSIM score) (see Figure 3 in supplementary material).

% CONCLUSION
\section{Discussion and Conclusion}
In this work, we propose \emph{HashEnc}, a solution based on grid-like local embeddings and replace SIREN in the NODF framework of \cite{nodf} to estimate the ODF field on high-resolution diffusion MRI scans. While SIREN suffers from over-smoothing high contrast regions, HashEnc learns better fine-grained structural features with significantly less training time, making it feasible for downstream tasks as reflected in the Feature Similarity Index (FSIM). We want to acknowledge that HashEnc is limited in its ability to adapt to different noise levels in the image. Our training image contains varying levels of noise across different regions, which HashEnc does not consider, as the number of multi-resolution grids is fixed for all regions and $\sigma_e^2$ is assumed to be spatially constant. We encourage further research to address this limitation in future studies.

% BIBLIOGRAPHY
\bibliographystyle{splncs04}
\bibliography{egbib}

\clearpage
\section{Supplementary Figures}

\begin{figure}[h]
    \centering
    \includegraphics[scale=0.37]{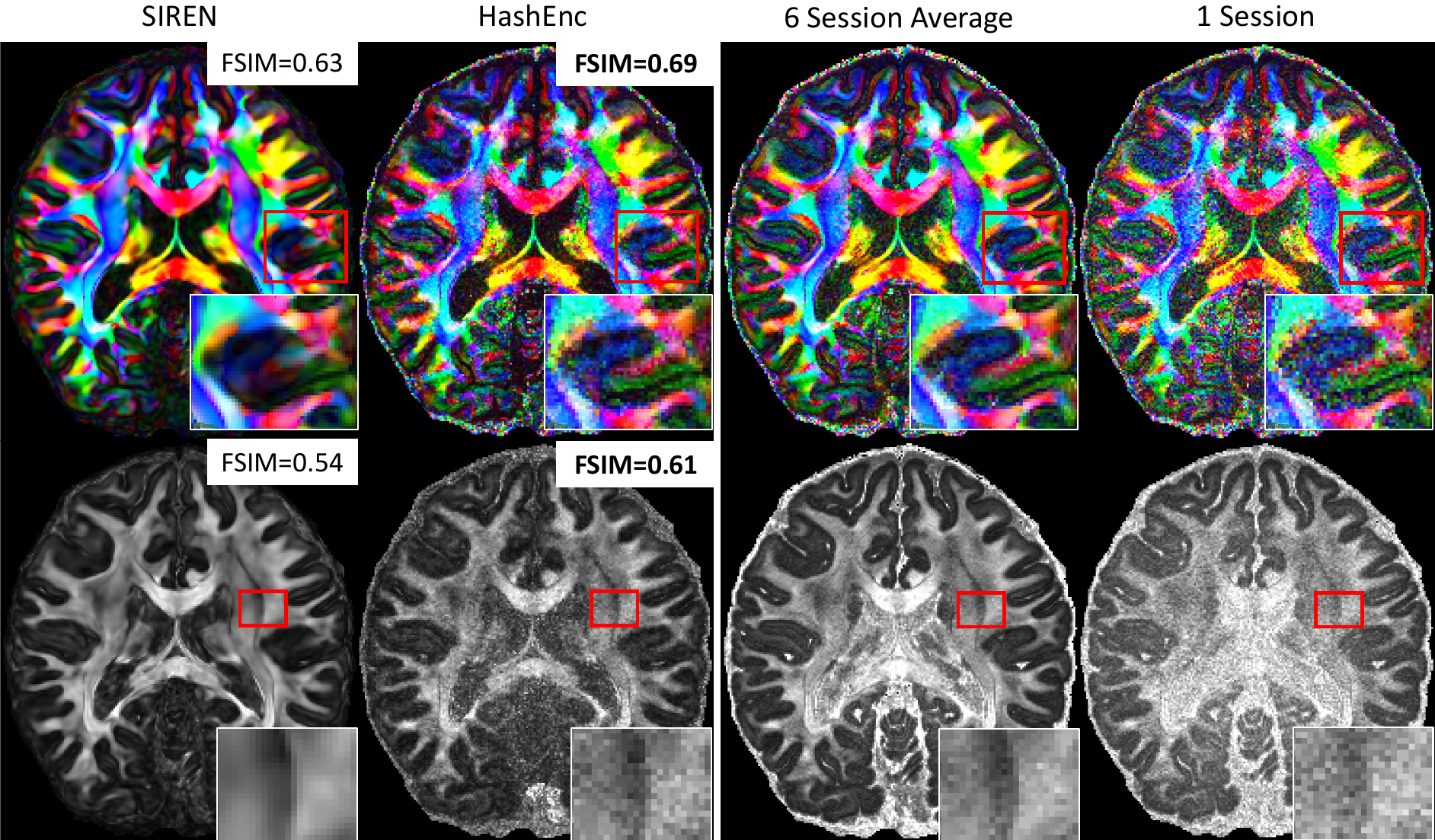}
    \caption{DTI and GFA images of an axial slice of SIREN and HashEnc methods trained on $M=70$ gradient directions. On the right side are the DTI and GFA images of the 6 session average and 1 session. For each of the SIREN and HashEnc images we report the FSIM score to the 6 session average. We also highlight a small section indicated by the red box to demonstrate the over-smoothing effect of SIREN in comparison to the other images. HashEnc shows a better structural similarity to the 6 session average, indicated both visually and by the higher FSIM score.}
    \label{fig:fsim}
\end{figure}

\begin{figure}[h]
    \centering
    \includegraphics[scale=0.35]{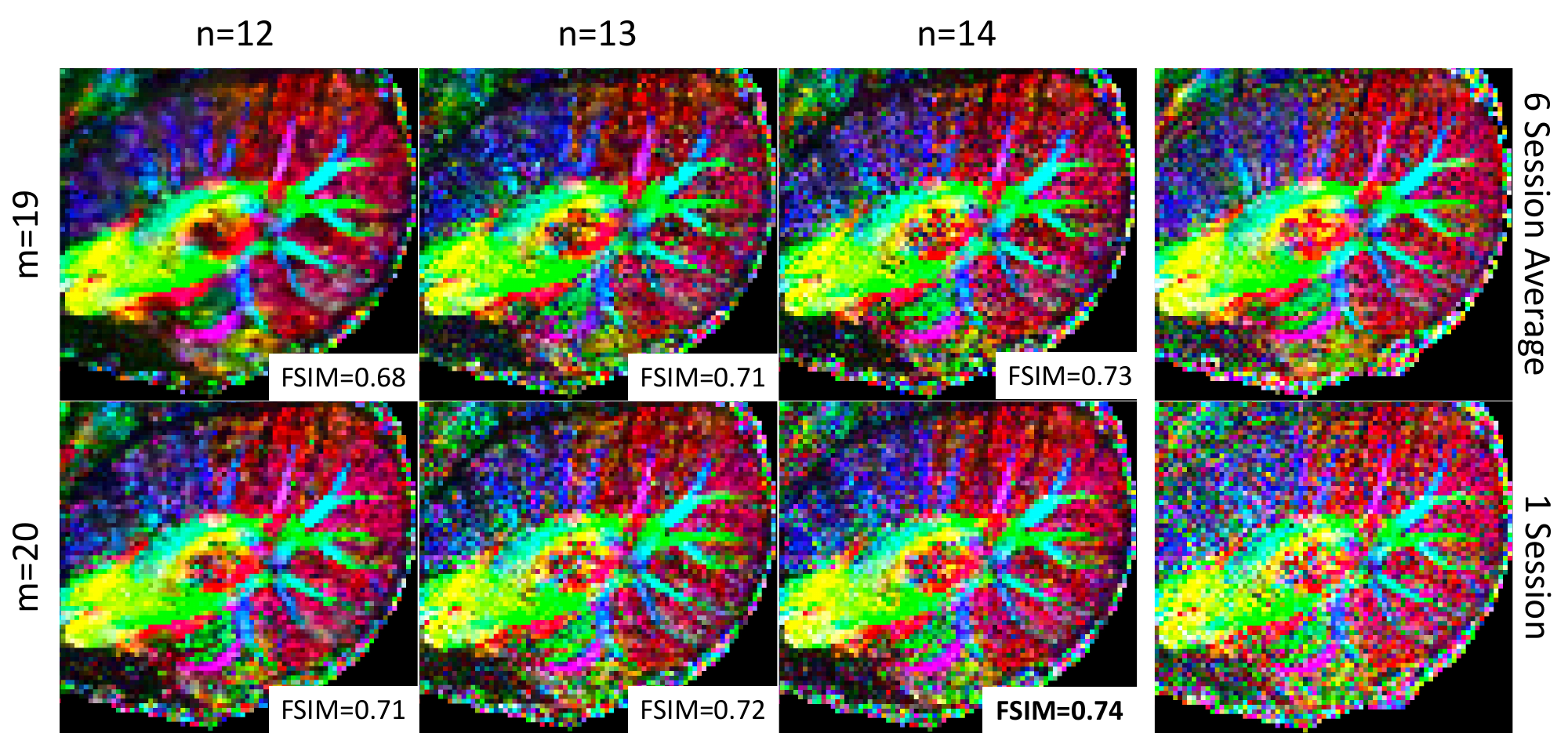}
    \includegraphics[scale=0.35]{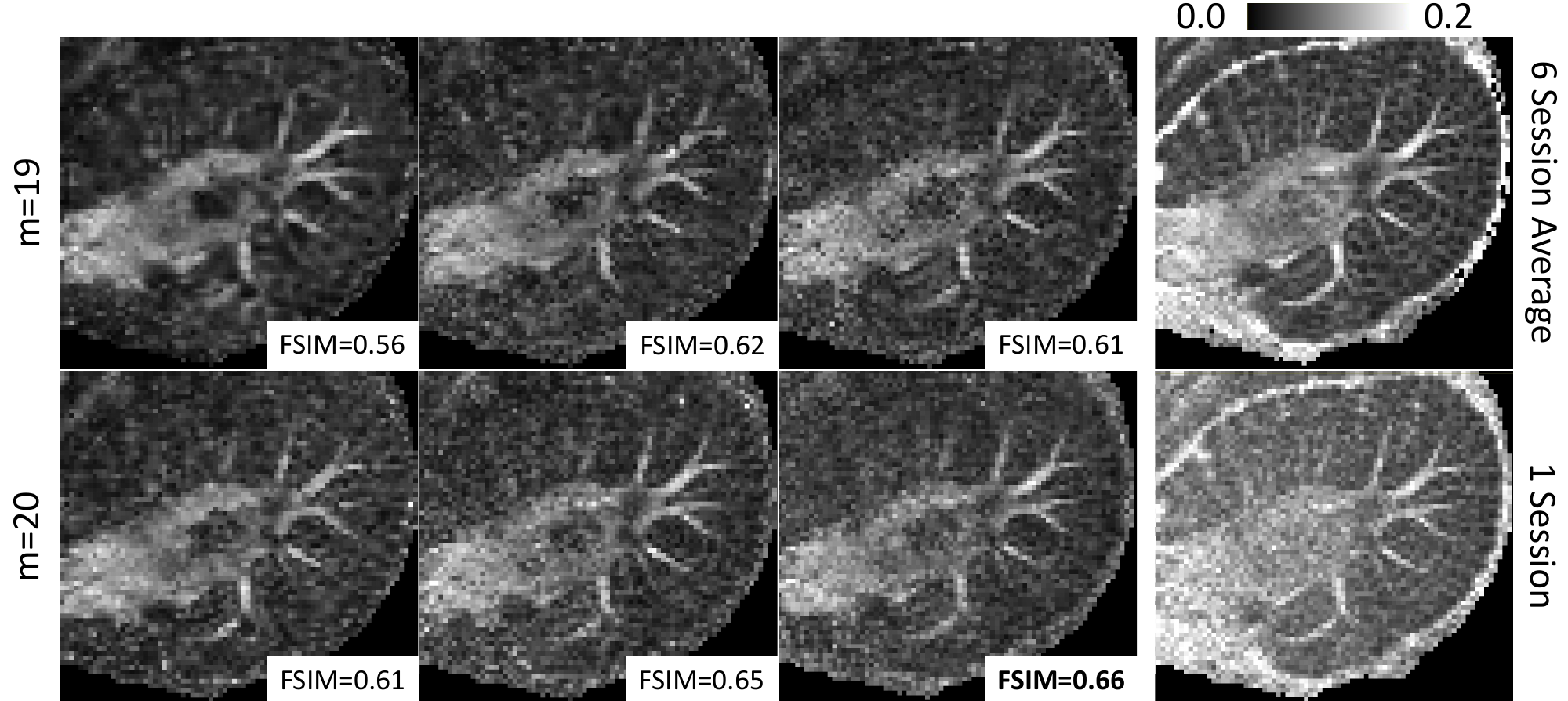}
    \caption{Cerebellum DTI and GFA images of HashEnc method with different resolutions levels $n$ and lookup table sizes $2^m$. Right are the 6 session average and 1 session images. We report the FSIM score of every image to the 6 session average on the bottom right corner. Based on the FSIM score, the network configuration of $n=14$ and $m=20$ shows the best structural similarity to the 6 session average.}
    \label{fig:ablation_hashenc_size}
\end{figure}

\begin{figure}[h]
    \centering
    \includegraphics[scale=0.36]{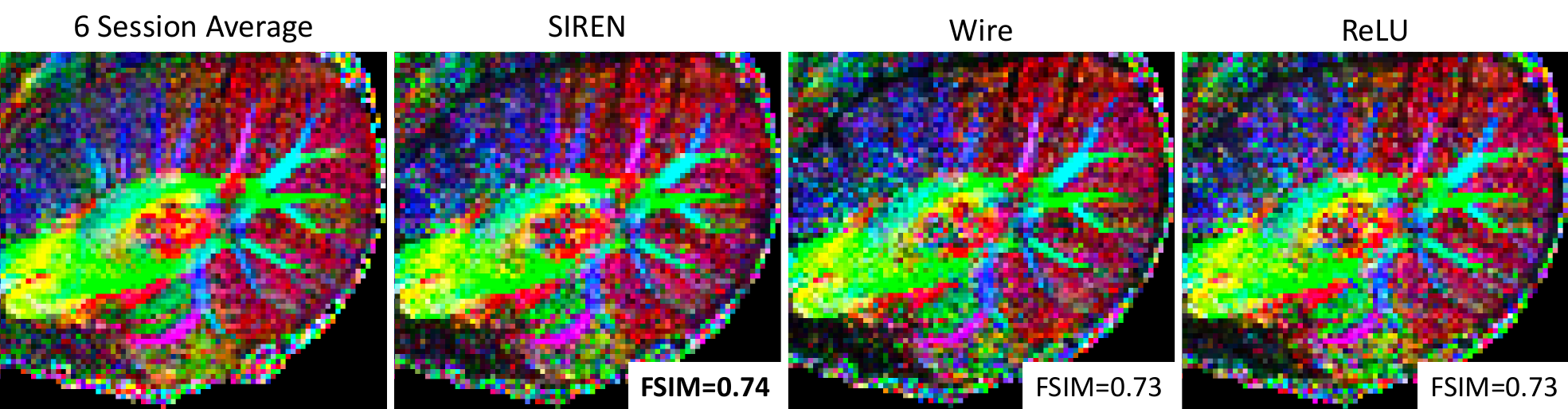}
    \caption{Cerebellum DTI images of HashEnc trained with different types of MLP heads (SIREN, Wire, and ReLU). We include the DTI image of the 6 session average on the left and the FSIM score of the rest of the images. All networks are trained with 14 resolution levels and a $2^{20}$ lookup table size on $M=70$ gradient directions. We see no significant difference in performance with different MLP heads.}
    \label{fig:ablation_mlp}
\end{figure}

\clearpage

\section{Additional Experiments}

In Section \ref{sec:tractography}, we present tractography images for SIREN and HashEnc. In Section \ref{sec:hyperparameter_tuning}, we provide a detailed discussion on the hyperparameter tuning for both HashEnc and SIREN.

\subsection{Tractography}
\label{sec:tractography}

\begin{figure}[h]
    \centering
    \includegraphics[scale=0.38]{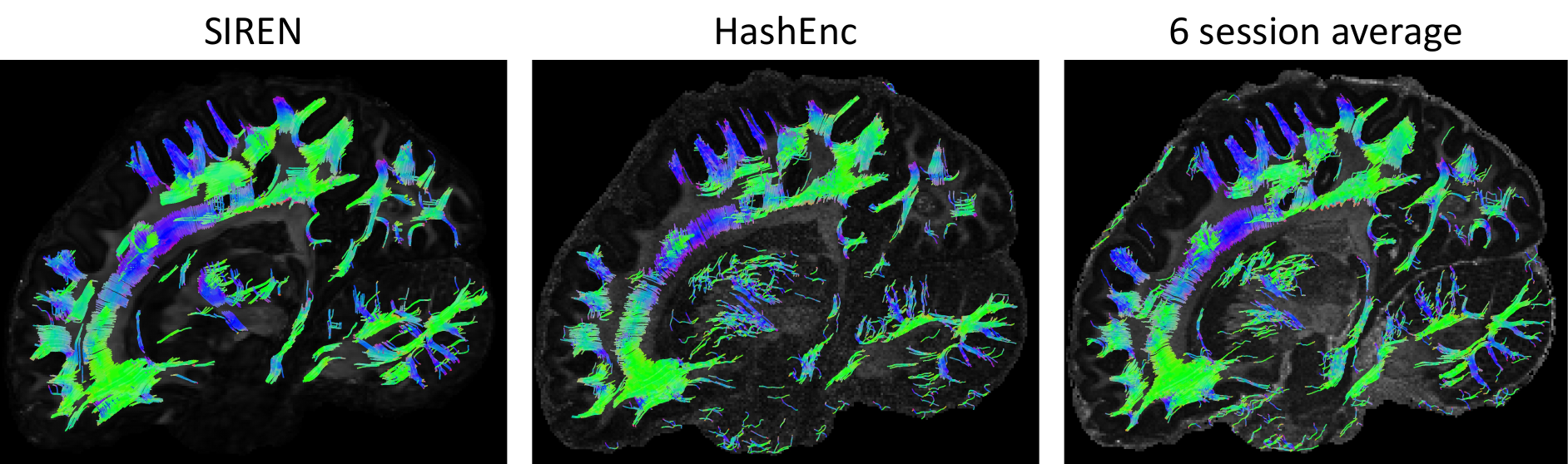}
    \caption{Sagittal slice showing the tractography results for HashEnc, SIREN, and the 6-session average ($M=70$). Tractography was obtained from deconvolved ODFs, GFA displayed in the background of each image. The sagittal slice corresponds to the same view presented in Fig. \ref{fig:time_comparison_updated} of the main text.}
    \label{fig:tractography}
\end{figure}

Figure \ref{fig:tractography} shows the tractography results for HashEnc and SIREN in comparison to the 6-session average. The tractography maps are obtained using the LocalTracking algorithm of the DIPY Python library applied to the estimated ODF fields. Peak detection was performed by first deconvolving the ODFs with constrained spherical deconvolution \cite{TOURNIER2007} and then calculating the local maxima of the deconvolved ODFs on a dense spherical mesh. Sample code implementing the full procedure is available on our GitHub \footnote{\url{https://github.com/MunzerDw/NODF-HashEnc/blob/main/evaluate.py\#L460}}. Relative to the 6-session average, the results indicate that HashEnc provides greater spatial coverage in certain regions, such as the center of the brain and the cerebellum, while SIREN demonstrates better performance in other areas, such as the recovery of tracks in the occipital region.

\subsection{Hyperparameter Tuning}
\label{sec:hyperparameter_tuning}

In supplemental experiments, we considered tuning various architectural hyperparameters for HashEnc, including the number of grid resolutions, the resolution of the starting grid, the scaling factor between grids, the MLP head size, the resolution of the highest grid, the size of the hashmap table, and $\lambda_c$. We evaluated performance both perceptually and in terms of the evaluation metrics on the GFA and DTI maps outlined in Section 4 of the main text. Through this fine-tuning, we are able to improve HashEnc's performance (see Fig. \ref{fig:time_comparison_updated} and Table \ref{tab:quant_results_updated}). The optimized hyperparameter values for HashEnc are as follows:

\begin{itemize}
    \item Increased MLP head to 3 layers of 128 neurons each. This increases the training time of HashEnc by approximately 6\%, which remains significantly lower than SIREN's training time.
    \item Decreased the number of grid layers to 4.
    \item Increased the number of features per grid embedding to 8.
    \item Increased base resolution to 80.
    \item Reduced grid level scaling factor to 1.13.
\end{itemize}

We note that these settings were found to be optimal for the specific dataset used in this study but would likely require adjustment for different images, particularly those with different resolutions and signal-to-noise ratios. 

\subsubsection{Regularization Penalty Strength.}
For the regularization strength $\lambda_c$, we experiment with different values ($1e^{-7}$, $1e^{-6}$, and $1e^{-5}$). For SIREN, higher $\lambda_c$ values result in some loss of detail, while lower values preserve more smaller details (particularly noticeable in the Cerebellum). For HashEnc, lower $\lambda_c$ values lead to increased noise capture, whereas higher values allow for better preservation of smaller details without introducing noise. The visual magnitude of these differences appears similar for both models, suggesting comparable sensitivity to changes in $\lambda_c$.

\subsubsection{Batch Size.}
\begin{figure}[!ht]
    \centering
    \includegraphics[scale=0.36]{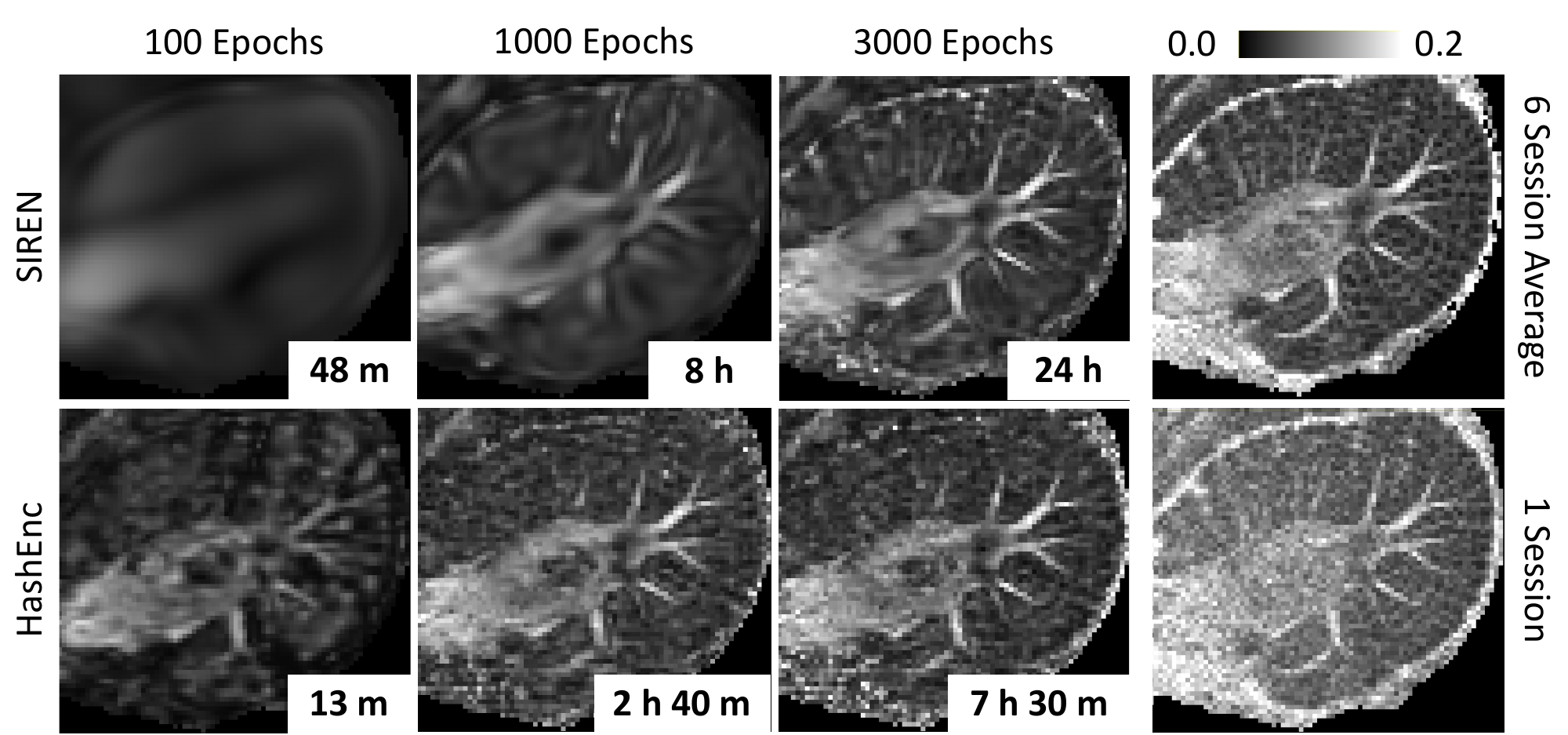}
    \caption{GFA reconstruction images from a sagittal cerebellum slice at different training times with $M=70$ gradient directions. The scale on the top right indicates the degree of anisotropy of water diffusion. Included are GFA of the training image (1 session) and the 6 session average for reference. Here, HashEnc has improved hyperparameters and both are trained with the new batch size. HashEnc fits a similarly detailed ODF field after significantly less training time compared to SIREN.}
    \label{fig:time_comparison_updated}
\end{figure}

Our experiments on hyperparameter tuning reveal that SIREN exhibits high sensitivity to batch size. With 3,985,192 voxels, we initially used a batch size of 60,862 (derived from 3,985,192 // 64), resulting in a final batch of 24. This configuration leads to SIREN being excessively smooth and volatility during training. Upon increasing the batch size to 65,536 ($2^{16}$), SIREN's performance improves significantly, matching that of HashEnc with optimized hyperparameters (see Fig. \ref{fig:time_comparison_updated} and Table \ref{tab:quant_results_updated}). Notably, HashEnc maintains consistent performance with the original batch size of 60,862. This discrepancy may be due to the fact that the small final batch of 24 only updates the relevant (local) hash embeddings in HashEnc, whereas in SIREN, all parameters are affected, highlighting the importance of careful batch size selection for global INR-based models. This relative insensitivity is a significant advantage for the HashEnc methods.

\begin{table}[!ht]
\caption{Median value of Feature Similarity Index (FSIM) to the 6 session average of all sagittal, axial and coronal slices for SIREN and HashEnc trained on $M=70$ gradient directions. Both network were carefully tuned for optimal hyperparameter selection. FSIM-GFA is calculated on gray scale GFA images, and FSIM-DTI is calculated on RGB DTI images. 1 means perfect similarity. SIREN slightly outperforms HashEnc on FSIM-GFA and both models perform equally on FSIM-DTI.}
\centering
% \resizebox{\textwidth}
% {!}{%
\setlength{\tabcolsep}{3pt}
\begin{tabular}{ccccc}
\toprule
Model & FSIM-GFA & FSIM-DTI \\ \midrule
SIREN    & \textbf{0.71} & \textbf{0.69} \\ 
HashEnc  & 0.67 & \textbf{0.69} \\  \bottomrule
\end{tabular}
\label{tab:quant_results_updated}
\end{table}
% %
% }

\end{document}